\documentclass[aps,prd,showpacs,preprintnumbers,twocolumn,groupedaddress,nofootinbib]{revtex4-1}
\pdfoutput=1

\usepackage[]{amsmath}
\usepackage[]{amssymb}
\usepackage[]{mathrsfs}
\usepackage[]{eucal}
\usepackage[]{tensor}
\usepackage[]{amsthm}
\usepackage[]{bm}

\usepackage[unicode]{hyperref}
\usepackage{xcolor}
\definecolor{pastgreen}{HTML}{669900}
\definecolor{pastblue}{HTML}{336699}
\definecolor{linkcol}{HTML}{663333}
\hypersetup{colorlinks,linkcolor={pastblue},citecolor={pastblue},urlcolor={pastblue}}

\newcommand{\pf}[1]{\mathbf{#1}}
\newcommand{\dd}{\partial}
\newcommand{\hdg}{\star} 
\newcommand{\df}{\mathrm{d}}
\newcommand{\w}{\wedge}

\newcommand{\nab}[1]{\nabla_{\!#1}}

\newcommand{\qqd}{\ , \quad}
\newcommand{\bc}{\begin{center}}
\newcommand{\ec}{\end{center}}
\newcommand{\be}{\begin{equation}}
\newcommand{\ee}{\end{equation}}

\newcommand{\defeq}{\mathrel{\mathop:}=}

\newcommand{\FF}{\mathcal{F}}
\newcommand{\GG}{\mathcal{G}}
\newcommand{\LL}{\mathscr{L}}

\newcommand{\tE}{\widetilde{E}}
\newcommand{\tB}{\widetilde{B}}

\usepackage{dsfont}
\newcommand{\nn}{\mathds{N}}

\newcommand{\cl}[1]{\overline{#1}}

\theoremstyle{plain} \newtheorem{tm}{Theorem}[]
\newcommand{\btm}{\begin{tm}}
\newcommand{\etm}{\end{tm}}

\begin{document}
\begin{flushleft}
\texttt{ZTF-EP-22-02}

\texttt{RBI-ThPhys-2022-23}
\end{flushleft}

\title{Constraints on singularity resolution by nonlinear electrodynamics}

\author{A. Bokuli\'c}
\email{abokulic@phy.hr}
\affiliation{Department of Physics, Faculty of Science, University of Zagreb, 10000 Zagreb, Croatia}
\author{T. Juri\'c}
\email{tjuric@irb.hr}
\affiliation{Rudjer Bo\v skovi\'c Institute, Bijeni\v cka cesta 54, HR-10002 Zagreb, Croatia}
\author{I. Smoli\'c}
\email{ismolic@phy.hr}
\affiliation{Department of Physics, Faculty of Science, University of Zagreb, 10000 Zagreb, Croatia}

\begin{abstract}
One of the long standing problems is a quest for regular black hole solutions, in which a resolution of the spacetime singularity has been achieved by some physically reasonable, classical field, before one resorts to the quantum gravity. The prospect of using nonlinear electromagnetic fields for this goal has been limited by the Bronnikov's no-go theorems, focused on Lagrangians depending on the electromagnetic invariant $F_{ab}F^{ab}$ only. We extend Bronnikov's results by taking into account Lagrangians that depend on both electromagnetic invariants, $F_{ab}F^{ab}$ and $F_{ab}\,{\hdg F^{ab}}$, and prove that the tension between the Lagrangian's Maxwellian weak field limit and boundedness of the curvature invariants persists in more general class of theories.
\end{abstract}

\maketitle

\section{Introduction} 

The electric field of a point charge, as well as its self-energy, are manifestly divergent in Maxwell's electrodynamics. Family of theories based on nonlinear modifications of Maxwell's Lagrangian, collectively called nonlinear electrodynamics (NLE), contains candidates that may resolve those singularities. For instance, phenomenological Born--Infeld Lagrangian \cite{Born34, BI34} puts the upper limit on the electric field strength, thus preventing it from diverging in the limit of short distances. Consequently, it also regularizes the energy of a point charge. Another prominent example is the effective Euler--Heisenberg Lagrangian \cite{HE36}, emanating from 1-loop QED calculation of the process of $\gamma\gamma\to\gamma\gamma$ scattering in the low energy limit. This theory removes the singularity in the energy of a point charge, but not necessarily in the electric field. Regularization of electrostatic quantities of a point charge is not achieved in the novel ModMax NLE theory \cite{BLST20}, based on a unique 1-parameter family of Lagrangians which respect both the conformal and the electromagnetic $SO(2)$ duality invariance, but can be achieved with further modifications of such Lagrangians \cite{Kruglov21,Kruglov22}. A broader class of NLE Lagrangians, those that satisfy the dominant energy condition and a number of technical assumptions, can regularize the electrostatic energy, as shown in \cite{TZ10}.

\smallskip

Singularities are also a ubiquitous feature of general relativity, manifested as some kind of curvature divergence or geodesic incompleteness. Since the first known exact black hole solutions were stationary and at least axially symmetric, it was unclear whether their singular behaviour is just an artefact of the artificially imposed symmetry. The analysis of the spacetime singularities culminated with the formulation of Hawking--Penrose singularity theorems \cite{TM1,TM2,TM3}. Assuming that certain energy conditions hold and requiring additional conditions on the causal structure of spacetime, the theorems imply the existence of incomplete geodesics \cite{SG15}, thus proving that singularities are not just ``by-products'' of the highly symmetric solutions. Geroch \cite{Ger} gave an example of geodesically complete spacetime, but which contains an incomplete nongeodesic timelike curve of bounded acceleration implying that even stricter regularity criteria are needed.

\smallskip

It is generally expected that quantum extensions of a classical theory should ``cure'' its singularities \cite{CDH21}. Before one invokes any of the proposed candidates for the quantum theory of gravitation (all of which have yet to be proven consistent, complete and experimentally verified), there are some other, less ambitious but quite important options. One is to replace the classical probe with the quantum one, and prove that a geodesically incomplete spacetime is in fact quantum complete \cite{Wald80, HM95}, or use the semiclassical backreaction as a mechanism to dress the singularity \cite{HS15, CFMZ16, *CFMZ17, Ty18, GJSS19}. The other is to inspect various generalizations of the Einstein--Maxwell theory, which is the venue we shall investigate in this paper.

\smallskip

Relying on the analogy with electromagnetism, it was hoped that NLE Lagrangian coupled to the gravitational sector could cure the spacetime singularities \cite{ABG98,ABG99}. This idea flourished after it was inferred that the \textit{ad hoc} proposed regular metric of Bardeen's black hole \cite{Bardeen68} can be obtained from NLE Lagrangian \cite{ABG00}. Bronnikov \cite{Bronnikov00} later established a general criterion under which a static, spherically symmetric solution of the Einstein--NLE field equations, with the NLE Lagrangian depending only on the invariant $F_{ab}F^{ab}$ and obeying the Maxwellian weak field limit, can have a regular center. The main conclusion is that the presence of electric charge prevents construction of a regular black hole solution (see, e.g., some examples in \cite{Hendi13}) and to this end one must rely only on magnetically charged solution. Indeed, electrically charged regular black holes constructed in \cite{FW, *Bronnikov17, GDGC, *Bronncom, PR} violate Maxwellian limit, while magnetically charged regular black holes in \cite{MTK, MA, AS19, KRUG1, *KRUG2, *KRUG3} do not. Dymnikova \cite{DY04} showed that by relaxing Bronnikov's conditions (precisely, discarding Maxwellian limit), it is possible to obtain regular electrically charged black hole solution with so-called ``de Sitter core'', de Sitter behaviour as $r\to 0$ (see also \cite{BCCH20,COS22}). Another evision of the Bronnikov's no-go theorem was proposed in \cite{BH}, based on a specific construction with core simulating a phase transition. Completely different approach, so-called double copy procedure, has been recently employed \cite{PT20, *MS22} for the construction of regular black holes via NLE fields.

\smallskip

As most of the NLE extensions of Maxwell's electromagnetism, emanating from some concrete quantum theory, have Lagrangians which depend on both electromagnetic invariants, $F_{ab}F^{ab}$ and $F_{ab}\,{\hdg F^{ab}}$, our main objective is to explore to which extent Bronnikov's results can be generalized. First of all we have to define what exactly do we mean by the \emph{regular} solution. Hereafter we will follow the classification and nomenclature of singularities as presented by Ellis and Schmidt in \cite{ES}. Our focus will be on scalar singularities, which occur if the spacetime is not further extendible and curvature scalars are not ``well behaved''. Namely, as curvature scalars are coordinate independent,  they must stay bounded in a regular spacetime.  Conversely, bounded curvature scalars do not guarantee the regularity of  spacetime, as there are geodesically incomplete spacetimes with vanishing curvature scalars \cite{Wald}. Nevertheless, scalar singularities carry enough information to formulate a no-go theorem since finding at least one diverging curvature invariant labels the spacetime as singular.

\smallskip

The paper is organised as follows. In Sec.~II we briefly summarize the basic aspects of NLE in the context of gravitational theory. Cornerstone of the argument, relation between the curvature and the electromagnetic invariants, is established in Sec.~III. In order to investigate all the invariants that may be obtained by contractions of arbitrary number of NLE energy momentum tensors, we express them in a closed form via spinor formalism. The central result, analysis of the regularity of spherically symmetric spacetimes sourced by NLE Lagrangians obeying Maxwellian limit, is elaborated in Sec.~IV. In Sec.~V we discuss ramifications of our theorems and comment on the remaining open questions.

\smallskip

\textit{Notation and conventions.} We use the ``mostly plus'' metric signature and natural system of units in which $G = c = 4\pi\epsilon_0 = 1$. For differential forms we use either abstract index notation or boldface letters. Hodge dual of a $p$-form $\bm{\omega}$ is defined as 
\be
(\hdg\bm{\omega})_{a_{p+1} \dots a_4} \defeq \frac{1}{p!} \, \omega_{a_1 \dots a_p} \tensor{\epsilon}{^{a_1 \dots a_p}_{a_{p+1} \dots a_4}} \, . 
\ee
Partial derivatives of the Lagrangian density $\LL(\FF,\GG)$ are denoted by $\LL_\FF \defeq \dd_\FF \LL$, $\LL_\GG \defeq \dd_\GG \LL$, $\LL_{\FF\GG} \defeq \partial_\GG \partial_\FF \LL$, and so on. For any rank-2 tensor $\tensor{X}{^a_b}$ and $n \in \nn$ we use shorthand notation 
\be
\tensor{(X^n)}{^a_b} \defeq \tensor{X}{^a_{c_1}} \tensor{X}{^{c_1}_{c_2}} \cdots \tensor{X}{^{c_{n-1}}_b} \, .
\ee

\section{An overview of NLE} 

With the electromagnetic field tensor $F_{ab}$ at disposal, we can construct two independent quadratic electromagnetic invariants, $\FF \defeq F_{ab}F^{ab}$ and $\GG \defeq F_{ab} \, {\hdg F^{ab}}$. In Maxwell's electrodynamics, Lagrangian density is given as $\LL^{\mathrm{(Max)}} = -\FF/4$, while NLE Lagrangian density $\LL(\FF,\GG)$ can generally depend on both invariants. In order to categorize NLE Lagrangians, we will use the following terminology: $\FF$-class consists of Lagrangians depending on invariant $F_{ab}F^{ab}$ only, while $\FF\GG$-class Lagrangians depend on both invariants. We do not consider terms which include covariant derivatives of $F_{ab}$ or nonminimal coupling to the gravitational sector. Thus, the total Lagrangian 4-form,
\be
\pf{L} = \frac{1}{16\pi} \Big(R - 2\Lambda + 4\LL(\FF,\GG)\Big) \, {\hdg 1} \, ,
\ee
consists of Einstein--Hilbert gravitational contribution, containing Ricci scalar $R$ and the cosmological constant $\Lambda$, and the electromagnetic part. We say that a NLE Lagrangian density $\LL$ obeys the Maxwellian weak field (MWF) limit if $\LL_\FF \to -1/4$ and $\LL_\GG \to 0$ as $(\FF,\GG) \to (0,0)$. 

\smallskip

A useful way of expressing the NLE energy-momentum tensor is to separate it into Maxwell's part and the trace part
\be\label{eq:TNLE}
T_{ab} = -4\LL_\FF \widetilde{T}_{ab} + \frac{1}{4} \, T g_{ab} \, ,
\ee
where the Maxwell's tensor $\widetilde{T}_{ab}$ and trace $T \defeq g^{ab} T_{ab}$ are, respectively, given by
\begin{align}
\widetilde{T}_{ab} = \frac{1}{4\pi} \left( F_{ac} \tensor{F}{_b^c} - \frac{1}{4} \, g_{ab} \, \FF \right) , \label{eq:TMax} \\
T = \frac{1}{\pi} \, (\LL - \LL_\FF \FF - \LL_\GG \GG) \, .
\end{align}
Introducing the auxiliary 2-form $\pf{Z}$ 
\be
\pf{Z} \defeq -4 \left( \LL_\FF \,\pf{F} + \LL_\GG \, {\hdg\pf{F}} \right) ,
\ee
generalized source-free NLE Maxwell's equations can be written as 
\be\label{eq:dFdZ}
\df\pf{F} = 0 \quad \textrm{and} \quad \df{\hdg\pf{Z}} = 0 \, .
\ee
Einstein's gravitational field equation sourced by the NLE energy-momentum tensor is
\be\label{eq:Einstein}
R_{ab} - \frac{1}{2} \, R \, g_{ab} + \Lambda g_{ab} = 8\pi T_{ab} \, .
\ee
For more comprehensive overview of NLE theories and their properties we refer reader to classic lectures by Pleba\'nski \cite{Plebanski70}, as well as some more recent papers \cite{Sorokin21,BJS21,BJS22}.

\section{A distillate of useful invariants} 

In order to examine the regularity of the spacetime, we will inspect the behaviour of curvature invariants that may be translated, via Einstein's gravitational field equation, into electromagnetic invariants. As on the gravitational side we have contractions of the Ricci tensor $R_{ab}$ on our disposal, the question is how many different invariants may be constructed by contractions of the energy-momentum tensor $T_{ab}$.

\smallskip

The evaluation of these contractions is most easily performed using spinor calculus \cite{PR1,Stewart}. Spinor space is endowed with the sympletic structure, antisymmetric nondegenerate spinor $\epsilon_{AB}$, and the electromagnetic field is represented with the symmetric spinor $\phi_{AB}$. Respecting the antisymmetry of the electromagnetic field tensor and its Hodge dual, their spinor counterparts can be written as
\begin{align}
F_{ABA'B'} & = \epsilon_{AB} \cl{\phi}_{A'B'} + \phi_{AB} \epsilon_{A'B'} \, , \\
{\hdg F}_{ABA'B'} & = i(\epsilon_{AB} \cl{\phi}_{A'B'} - \phi_{AB} \epsilon_{A'B'}) \, .
\end{align}
Using this decomposition, it is straightforward to express the electromagnetic invariants 
\begin{align}
\FF & = 2(\phi^{AB}\phi_{AB} + \cl{\phi}^{A'B'}\cl{\phi}_{A'B'}) \, , \\
\GG & = -2i(\phi^{AB}\phi_{AB} - \cl{\phi}^{A'B'}\cl{\phi}_{A'B'}) \, .
\end{align}
Maxwell's energy momentum tensor (\ref{eq:TMax}) in spinor form is given by
\be
\widetilde{T}_{ABA'B'} = \frac{1}{2\pi} \, \phi_{AB} \cl{\phi}_{A'B'} \, ,
\ee
where we have used
\be
\phi_{AC} \tensor{\phi}{_B^C} = \frac{1}{2} \, (\phi_{CD} \phi^{CD}) \epsilon_{AB} \, .
\ee
The trace of the odd number of Maxwell's energy-momentum tensors vanishes as it is proportional to the contraction of the symmetric spinor $\phi_{AB}$ with the antisymmetric spinor $\epsilon_{AB}$,
\be
\tensor{(\widetilde{T}^{2n+1})}{^a_a} = 0 \, .
\ee
The trace of the even number of Maxwell's energy-momentum tensors reduces to 
\be
(4\pi)^{2n} \tensor{(\widetilde{T}^{2n})}{^a_a} = \frac{1}{4^{2n-1}} \, (\FF^2 + \GG^2)^n \, .
\ee
Taking into account the expression above, we can easily evaluate the trace of two NLE energy-momentum tensors (\ref{eq:TNLE})
\be\label{eq:TT}
4\pi^2 \tensor{T}{^a_b} \tensor{T}{^b_a} = \pi^2 T^2 + \LL_\FF^2 (\FF^2 + \GG^2) \, .
\ee
In fact, using the binomial formula and Eq.~(\ref{eq:TNLE}), it is not difficult to generalize Eq.~(\ref{eq:TT}) for an arbitrary number of contracted energy-momentum tensors,
\be
\tensor{(T^n)}{^a_a} = 4 (T/4)^n + \sum_{k=1}^n \binom{n}{k} 4^{2k-n} (-\LL_\FF)^k T^{n-k} \tensor{(\widetilde{T}^k)}{^a_a} \, .
\ee
The $T^n$ term above is, just for clarity, written separately. As can be seen, all these contractions are always reduced to combinations of the two basic ones, the trace $T$ and $\LL_\FF^2 (\FF^2 + \GG^2)$, upon which we shall base our further discussion.

\smallskip

To summarize, Einstein's field equation (\ref{eq:Einstein}) provides us with with the relation between the curvature and electromagnetic invariants,
\begin{align}
R - 4\Lambda & = -8\pi T \, , \label{eq:RTRRTT1} \\
R_{ab} R^{ab} + 2\Lambda(2\Lambda - R) & = (8\pi)^2 \, T_{ab} T^{ab} \, , \label{eq:RTRRTT2}
\end{align}
so that the boundedness of Ricci scalar $R$ and Ricci squared $R_{ab}R^{ab}$ translates to the boundedness of the energy-momentum invariants $T_{ab}T^{ab}$ and trace $T$. Cosmological constant $\Lambda$ is included for the sake of generality, but its role in the following arguments is mostly passive, as we do not rely on the asymptotic properties of the spacetime.

\section{Constraints} 

Our main analysis will be, for simplicity, focused on the static, spherically symmetric spacetimes. Namely, we demand from a candidate theory to achieve regularization of an \emph{arbitrary} black hole solution, without aid of, for example, additional angular momentum. Static, spherically symmetric metric can be put in the form \cite{Wald}
\be
\df s^2 = -\alpha(r) \, \df t^2 + \beta(r) \, \df r^2 + r^2 \left( \df\theta^2 + \sin^2\theta \, \df\varphi^2 \right) \, ,
\ee
given that $\nab{a} r \ne 0$. We assume that the radial coordinate $r$ attains its minimum $r = 0$ at a point referred to as a \emph{center}, which will be assumed to be regular in a sense defined below. Specific cases not covered by this geometric setting, such as a wormhole solutions (in which $r$ attains some minimal value $r_* > 0$) or solutions with a ``horn'' (infinitely long tube of finite radius), can be set aside due to Bronnikov's theorem 2 \cite{Bronnikov00}, whose assumption $\tensor{T}{^t_t} = \tensor{T}{^r_r}$ is satisfied in our context. Furthermore, we introduce, for convenience, the abbreviation $w(r) \defeq \sqrt{\alpha(r)\beta(r)}$ and assume that on some punctured neighbourhood of the center $r = 0$, e.g.~points with $0 < r < r_w$ for some $r_w > 0$, $w(r)$ has no zeros. In other words, we assume that at least in some neighbourhood of the center there are no horizons. We note in passing that the condition $\tensor{T}{^t_t} = \tensor{T}{^r_r}$ is also sufficient \cite{Jacobson07}, at least with the Einstein field equation, to simply take $w(r) = 1$ without loss of generality, but we shall leave the function $w$ undetermined for the sake of possible generalizations beyond the Einstein--Hilbert theory.

\smallskip

When we say that some scalar $\psi(r)$ is \emph{bounded as $r \to 0$}, we assume that there is a real constant $M > 0$ and a radius $r_0 > 0$, such that $|\psi(r)| \le M$ for all $0 < r < r_0$. One must bear in mind that such criterion of boundedness of a scalar is quite mild: We \emph{do not} assume a priori that the limit $\lim_{r\to 0} \psi(r)$ necessarily exists (e.g.~$\psi$ could widely oscillate, something like $\psi(r) \sim \sin(1/r)$, as we approach the center). Thus, even if a certain spacetime passes this low-bar test for a number of invariants, any of them may still be rather ill-behaved in a neighbourhood of the center.

\smallskip

We shall introduce two auxiliary 1-forms, electric $E_a \defeq -k^b F_{ba}$ and magnetic $B_a \defeq k^b {\hdg F}_{ba}$, defined with respect to the Killing vector field $k = \dd/\dd t$. The electromagnetic 2-form, which inherits the spacetime symmetries,\footnote{General problem of symmetry inheritance for NLE fields is discussed in \cite{BGS17}.} is given by
\begin{align}
\pf{F} & = -E_r(r) \, \df t \w \df r - B_r(r) \, {\hdg (\df t \w \df r)} \\
 & = -E_r(r) \, \df t \w \df r + \frac{B_r(r)}{w(r)} \, r^2 \sin\theta \, \df\theta \w \df\varphi \, ,
\end{align}
and its corresponding Hodge dual by
\be
{\hdg\pf{F}} = \frac{E_r(r)}{w(r)} \, r^2 \sin\theta \, \df\theta \w \df\varphi + B_r(r) \, \df t \w \df r \, .
\ee
Furthermore, just for convenience, we shall introduce rescaled electric and magnetic 1-forms
\be
\tE_a \defeq \frac{E_a}{w} \qqd \tB_a \defeq \frac{B_a}{w} \, .
\ee
Two corresponding electromagnetic invariants are then given by
\be
\FF = 2 (\tB_r^2 - \tE_r^2) \quad \textrm{and} \quad \GG = 4 \tE_r \tB_r \, .
\ee
Now, NLE Maxwell's equations (\ref{eq:dFdZ}) immediately imply that $\tB_r r^2$ and $(\LL_\FF \tE_r - \LL_\GG \tB_r) r^2$ are constants, which can be fixed using definitions of the electric charge $Q$ and the magnetic charge $P$ given by the Komar integrals, evaluated over a sphere $\mathcal{S}$,
\be
Q \defeq \frac{1}{4\pi} \oint_\mathcal{S} {\hdg\pf{Z}} \quad \textrm{and} \quad P \defeq \frac{1}{4\pi} \oint_\mathcal{S} {\pf{F}} \, .
\ee
Choice of the sphere $\mathcal{S}$ is essentially irrelevant (up to technical obstacles, such as a question of proper coordinate system at the event horizon), as we are looking at source-free Maxwell's equations. This gives us finally
\begin{align}
\tB_r & = \frac{P}{r^2} \, , \label{eq:Max1} \\
\LL_\FF \tE_r - \LL_\GG \tB_r & = -\frac{Q}{4r^2} \, . \label{eq:Max2}
\end{align}
Now we turn to the analysis of the constraints on singularity resolution by nonlinear modifications of the Maxwell's electromagnetism.

\smallskip

Basic strategy for the main results is to assume that both $R$ and $R_{ab} R^{ab}$ are bounded as $r \to 0$, implying via Einstein's gravitational field equation, as shown in Eqs.~(\ref{eq:RTRRTT1})--(\ref{eq:RTRRTT2}), that the same has to hold for $T$ and $T_{ab} T^{ab}$, which in turn implies, via Eq.~(\ref{eq:TT}), the boundedness of both $\LL_\FF \FF$ and $\LL_\FF \GG$ as $r \to 0$. As we shall show below, this assumption in many important cases cannot be consistent with WMF limit of a NLE Lagrangian. Again, it is important to stress that we impose only a mild regularity condition, namely boundedness of just two curvature scalars, $R$ and $R_{ab} R^{ab}$, which by itself does not prevent any other independent curvature scalar, such as the Kretschmann scalar $R_{abcd} R^{abcd}$, to diverge. However, even such seemingly benign assumption will be enough to produce strong constraints.

\subsection{Electric case} 

Given that magnetic monopoles have not yet been discovered\footnote{Quite intriguingly, magnetic charge on black holes may be bounded via its shadow \cite{AKVM20}.}, the most important case is the one in which a black hole bears only electric charge. Here we have a strong generalization of the Bronnikov's result \cite{Bronnikov00}.

\btm\label{tm:1}
Suppose that the spacetime is a static, spherically symmetric solution of the Einstein--NLE field equations with $\FF\GG$-class NLE Lagrangian obeying Maxwellian weak field limit.  Then, in the electrically charged case, that is $P=0$ and $Q\ne 0$, Ricci scalar $R$ and Ricci squared $R_{ab}R^{ab}$ cannot both remain bounded as $r\to 0$.
\etm

Note that the Theorem \ref{tm:1} automatically applies to all $\FF$-class NLE Lagrangians.

\smallskip

\emph{Proof of Theorem \ref{tm:1}}. Absence of the magnetic charge, $P = 0$, immediately implies $\tB_r = 0$, allowing us to rewrite Maxwell's equation (\ref{eq:Max2}), after squaring and multiplication by $\FF$, as 
\be
\frac{\FF}{r^3} = -\frac{8}{Q^2} \, (\FF \LL_\FF)^2 r \, .
\ee
If both $R$ and $R_{ab}R^{ab}$ are bounded as $r\to 0$, then the same holds for $\FF \LL_\FF$, which implies that $\FF = o(r^3)$ as $r \to 0$. Also, as $\tB_r = 0$, the other electromagnetic invariant $\GG$ is identically zero. Finally, as $\LL_\FF^2 = -Q^2/(8\FF r^4)$, we can deduce that $\LL_\FF$ is unbounded as $r\to 0$, in direct contradiction with the assumed MWF limit. Note that the contradiction with MWF limit is manifest due to a fortunate occurrence: the $r\to 0$ limit coincides with the weak field limit in which both $\FF$ and $\GG$ approach zero. \qed

\smallskip

The obtained result comes as no surprise, as we know that electrically charged Born--Infeld \cite{GSP84, SGP87, FK03, Dey04} and Euler--Heisenberg \cite{YT00, RWX13} black holes are not regular (cf.~\cite{DARG10} for an in-depth analysis).

\subsection{Dyonic case} 

In the dyonic case, we cannot directly utilise the same procedure, since the weak field limit is not necessarily captured as we approach the center. Namely, using Maxwell's equation (\ref{eq:Max1}), electromagnetic invariant $\FF$ may be related to the other invariant $\GG$ via
\be\label{eq:FPPG}
\FF = 2 \left( \frac{P^2}{r^4}  - \frac{r^4}{16 P^2}\,\GG^2 \right) \, .
\ee
Here it is manifest that the origin of the $\FF$-$\GG$ plane is unattainable as $r \to 0$, which essentially takes away the opportunity to directly test the MWF limit. However, this very relation may be used for slightly different approach: Given that one proves that both $\FF$ and $\GG$ should, under some assumptions, remain bounded as $r \to 0$, we immediately have a contradiction.

\smallskip

Furthermore, using the definition of the invariant $\GG$ and Maxwell's equations (\ref{eq:Max1})--(\ref{eq:Max2}), we obtain
\begin{align}
\LL_\FF \tE_r & = \LL_\FF \GG \, \frac{r^2}{4P} \nonumber \\
 & = \LL_\GG \tB_r - \frac{Q}{4r^2} = \frac{1}{r^2} \left( \LL_\GG P - \frac{Q}{4} \right) ,
\end{align}
that is
\be\label{eq:LGLFG}
\frac{1}{r^3} \left( \LL_\GG P - \frac{Q}{4} \right) = \frac{\LL_\FF \GG}{4P} \, r \, .
\ee
From here, given that $\LL_\FF \GG$ remains bounded, it follows that
\be\label{eq:LGQP}
\LL_\GG = \frac{Q}{4P} + o(r^3) \quad \textrm{as} \quad r \to 0 \, .
\ee
This is another bounded invariant, particularly useful for the dyonic case. First we revisit Bronnikov's result \cite{Bronnikov00} with a slightly different proof.

\btm\label{tm:2}
Suppose that the spacetime is a static, spherically symmetric solution of the Einstein--NLE field equations with the $\FF$-class NLE Lagrangian. Then, in the dyonic case, that is $P\ne 0$ and $Q\ne 0$, Ricci scalar $R$ and Ricci squared $R_{ab}R^{ab}$ cannot both remain bounded as $r\to 0$.
\etm

\emph{Proof of Theorem \ref{tm:2}}. Let us assume that both $R$ and $R_{ab}R^{ab}$ are bounded as $r\to 0$, so that the same holds for $\LL_\FF \FF$ and $\LL_\FF \GG$. In the $\FF$-class case Maxwell's equations may be written as
\be
\LL_\FF \GG \, \frac{r^2}{4P} = \LL_\FF \tE_r = -\frac{Q}{4r^2}
\ee
which, given that by assumption $\LL_\FF \GG$ should remain bounded, immediately leads to a contradiction as $r \to 0$. \qed

\medskip

Note that the Theorem \ref{tm:2} relies only partly on MWF limit: We have identically $\LL_\GG = 0$, while we do not need to invoke that $\LL_\FF \to -1/4$ as $(\FF,\GG) \to (0,0)$.

\smallskip

We do not see how to generalize this result to \emph{all} NLE theories with $\FF\GG$-class NLE Lagrangians, so in order to make progress we shall focus on some special classes of NLE theories. First, without loss of generality, any $\FF\GG$-class NLE Lagrangian may be conveniently written as
\be\label{eq:NLEh}
\LL = -\frac{1}{4} \, \FF + h(\FF,\GG) \, ,
\ee
with some $C^1$-class function $h$. Two particular subclasses of NLE theories admit an easy generalization of constraints, first of which holds both for solutions with $Q \ne 0$ and $Q = 0$.

\btm\label{tm:3}
Suppose that the spacetime is a static, spherically symmetric solution of the Einstein--NLE field equations with the NLE Lagrangian (\ref{eq:NLEh}), such that $h = h(\GG)$. Then, given that $P\ne 0$, Ricci scalar $R$ and Ricci squared $R_{ab}R^{ab}$ cannot both remain bounded as $r\to 0$.
\etm

\smallskip

\emph{Proof of Theorem \ref{tm:3}}. As $\LL_\FF = -1/4$ identically, boundedness of $\LL_\FF\FF$ and $\LL_\FF\GG$ immediately implies boundedness of $\FF$ and $\GG$ as $r \to 0$, which in turn leads to a contradiction. Note that this part of the theorem, just as the Theorem \ref{tm:2}, relies only partly on MWF limit: we have identically $\LL_\FF = -1/4$, while we do not need to invoke that $\LL_\GG \to 0$ as $(\FF,\GG) \to (0,0)$. \qed

\btm\label{tm:4}
Suppose that the spacetime is a static, spherically symmetric solution of the Einstein--NLE field equations with the NLE Lagrangian (\ref{eq:NLEh}), such that $h(\FF,\GG) = a\FF^s \GG^u$, with a real constant $a \ne 0$ and integers $s,u \ge 1$. Then, in the dyonic case, that is $P\ne 0$ and $Q\ne 0$, Ricci scalar $R$ and Ricci squared $R_{ab}R^{ab}$ cannot both remain bounded as $r\to 0$.
\etm

\emph{Proof of Theorem \ref{tm:4}}. As for this theory
\begin{align}
\pi T & = (1-s-u) h \, , \\
\LL_\FF \FF & = -\frac{1}{4} \, \FF + s h \, ,
\end{align}
it follows that boundedness of $T$ and $\FF \LL_\FF$ imply boundedness of $h$ and $\FF$ as $r \to 0$. Furthermore, using
\be
\LL_\GG \GG = u h \, ,
\ee
and (\ref{eq:LGQP}), it follows that $\GG$ is bounded as $r \to 0$, which immediately leads to a contradiction. \qed

\medskip

Furthermore, a prominent family of theories are those with $h$ which is simply a quadratic polynomial, appearing in low field limits of quantum corrections to classical Maxwell's electromagnetism.

\btm\label{tm:5}
Suppose that the spacetime is a static, spherically symmetric solution of the Einstein--NLE field equations with the NLE Lagrangian (\ref{eq:NLEh}), such that $h(\FF,\GG) = a\FF^2 + b\FF\GG + c\GG^2$, where $a$, $b$ and $c$ are real constants. Then, in the dyonic case, that is $P\ne 0$ and $Q\ne 0$, Ricci scalar $R$ and Ricci squared $R_{ab}R^{ab}$ cannot both remain bounded as $r\to 0$.
\etm

\emph{Proof of Theorem \ref{tm:5}}. Due to the simplicity of the Lagrangian, evaluation of the derivatives $\LL_\FF$ and $\LL_\GG$ translates into a linear system for $\FF$ and $\GG$,
\begin{align}
\LL_\FF + \frac{1}{4} = 2a\FF + b\GG \, , \label{eq:qLF} \\
\LL_\GG = b\FF + 2c\GG \, . \label{eq:qLG}
\end{align}
Furthermore, from (\ref{eq:LGQP}) we have
\be
(b\FF + 2c\GG) \LL_\FF = \left( \frac{Q}{4P} + o(r^3) \right) \LL_\FF \, . 
\ee
Thus, given that $\FF \LL_\FF$ and $\GG \LL_\FF$ remain bounded as $r \to 0$, this has to hold also for $\LL_\FF$ itself.

\smallskip

Now we have to distinguish two subcases, according to the determinant of the linear system above, $\Delta = 4ac - b^2$. In the nondegenerate case, that is $\Delta \ne 0$, boundedness of $\LL_\FF$ and $\LL_\GG$ implies that invariants $\FF$ and $\GG$ are bounded as $r \to 0$, which leads to contradiction. In the degenerate case $\Delta = 0$, we need to carefully examine further subcases. If $c = 0$, then $b = 0$, and we are back at the $\FF$-class Lagrangian, covered by the Theorem \ref{tm:2}. If $a = 0$, then $b = 0$, and we are back at the Theorem \ref{tm:3}. Thus, let us assume that $a \ne 0 \ne c$. If we multiply both sides of
\be
\LL_\FF = -\frac{1}{4} + \frac{2a}{b} \, \LL_\GG
\ee
by $\FF$ and use Eq.~(\ref{eq:LGQP}), we get
\be
\FF\LL_\FF = \left( -\frac{1}{4} + \frac{aQ}{2bP} + o(r^3) \right) \FF \quad \textrm{as} \quad r \to 0 \, .
\ee
Here we have another two subcases. If $2aQ \ne bP$, then we may deduce that $\FF$ is bounded as $r \to 0$. Consequently, from Eq.~(\ref{eq:qLG}) we also see that $\GG$ is bounded as $r \to 0$, leading to a contradiction. In the remaining subcase when $2aQ = bP$, Eqs.~(\ref{eq:qLF})--(\ref{eq:qLG}) imply
\be
P\LL_\GG - Q\LL_\FF = \frac{Q}{4} \, , 
\ee
so that, via Eq.~(\ref{eq:LGLFG}),
\be
\GG = \frac{4P}{\LL_\FF r^4} \left( P\LL_\GG - \frac{Q}{4} \right) = \frac{4QP}{r^4}
\ee
and, using Eq.~(\ref{eq:FPPG}),
\be
\FF = \frac{2}{r^4} \, (P^2 - Q^2) \, ,
\ee
which, inserted into Eq.~(\ref{eq:qLG}), gives us
\be
\LL_\GG = \frac{2b}{r^4} \, (Q^2 + P^2) \, .
\ee
This, again, leads to a contradiction, as the right-hand side is manifestly unbounded as $r \to 0$, whereas the left-hand side should be bounded according to Eq.~(\ref{eq:LGQP}). \qed

\medskip

Finally, we turn to two distinguished NLE theories not covered by the Theorems \ref{tm:2}--\ref{tm:5}. Born--Infeld theory \cite{BI34} is defined with the 1-parameter $\FF\GG$-class Lagrangian
\be\label{L:BI}
\LL^{\mathrm{(BI)}} = b^2 \left( 1 - \sqrt{1 + \frac{\FF}{2b^2} - \frac{\GG^2}{16b^4}} \right) \, ,
\ee
where the real parameter $b > 0$ is physically related to the upper bound of the point charge electric field. It is straightforward to check that Born--Infeld Lagrangian respects MWF limit. ModMax NLE Lagrangian \cite{BLST20}
\be\label{L:MM}
\LL^{\mathrm{(MM)}} = \frac{1}{4} \left( -\FF\cosh\gamma +\sqrt{\FF^2 + \GG^2} \, \sinh\gamma \right) ,
\ee
is defined with the real parameter $\gamma$, but does not have well-defined partial derivatives $\LL^{\mathrm{(MM)}}_\FF$ and $\LL^{\mathrm{(MM)}}_\GG$ in the $(\FF,\GG) \to (0,0)$ limit (thus, strictly speaking, ModMax Lagrangian does not respect the MWF limit, as defined in this paper). We note in passing that the ModMax theory also appears in recent investigations of so-called $T\overline{T}$ deformations \cite{CINT18,BAVYM22,FSTM22,*FSSTM22}.

\smallskip

\btm\label{tm:6}
Suppose that the spacetime is a static, spherically symmetric solution of the Einstein--NLE field equations with the Born--Infeld (\ref{L:BI}) or ModMax (\ref{L:MM}) NLE Lagrangian. Then, given that $P\ne 0$, Ricci scalar $R$ and Ricci squared $R_{ab}R^{ab}$ cannot both remain bounded as $r\to 0$.
\etm

\emph{Proof of Theorem \ref{tm:6}}.

\smallskip

(a) Born--Infeld theory. Let us first assume that $Q \ne 0$. By looking at $\FF\LL_\FF$ and $\LL_\GG$ as a system for $\FF$ and $\GG$ we get\footnote{Sign ambiguity at the $W$ term appears as we cannot uniquely determine invariant $\GG$ from $\FF$, $\LL_\FF$ and $\LL_\GG$ alone, but is not relevant for our result.}
\be
\GG = -16\LL_\GG \frac{\FF\LL_\FF \pm W}{1 + 16\LL_\GG^2}
\ee
with
\be
W \defeq \sqrt{(\FF\LL_\FF)^2 + b^4(1 + 16\LL_\GG^2)} \, .
\ee
Again, if we assume that $R_{ab} R^{ab}$ and $R$ are bounded, then we immediately conclude that $\GG$ is bounded as $r \to 0$. Then, using
\be
4b^2 \LL_\GG \FF = -(\LL_\FF \FF) \GG \, ,
\ee
it follows that $\FF$ is also bounded as $r \to 0$, leading to a contradiction. If $Q = 0$, we can use a slightly different strategy: Maxwell's equation (\ref{eq:Max2}) leads to
\be
\left( 1 + \frac{P^2}{b^2 r^4} \right) \LL_\FF \tE_r = 0 \, ,
\ee
which implies $\LL_\FF \tE_r = 0$ for all points $r > 0$. Now, as
\be
\LL_\GG = -\frac{P}{(br)^2} \, \LL_\FF \tE_r \, ,
\ee
it follows that $\LL_\GG = 0$ and, given that both the trace $T$ and $\FF\LL_\FF$ should remain bounded, the same should hold for the Lagrangian itself. Thus, $\LL_\FF$ has no zeros for $r > 0$ and we may infer that $\tE_r = 0$, so that $\GG = 0$, $\FF = 2P^2/r^4$ and
\be
\FF \LL_\FF = -\frac{bP^2}{2r^2 \sqrt{P^2 + b^2 r^4}} \, ,
\ee
which is manifestly not bounded as $r \to 0$.

\smallskip

(b) ModMax theory. In the dyonic case we have (cf.~also \cite{FAGMLM20})
\be
\tE_r = \frac{Qe^{-\gamma}}{r^2} \qqd \tB_r = \frac{P}{r^2} \ ,
\ee
and, by direct evaluation,
\begin{align}
\FF\LL_\FF & = \frac{Q^2 + P^2}{2e^\gamma r^4} \, \frac{Q^2 - P^2 e^{2\gamma}}{Q^2 + P^2 e^{2\gamma}} \, , \\
\GG\LL_\FF & = -\frac{PQ}{r^4} \, \frac{Q^2 + P^2}{Q^2 + P^2 e^{2\gamma}} \, ,
\end{align}
which cannot both remain bounded as $r \to 0$, unless $Q = 0 = P$. \qed

\smallskip

In other words, both Born--Infeld and ModMax generalizations of the dyonic Reissner--Nordstr\"om solution still have unbounded curvature invariants at their center and in this sense cannot be considered as regularized black holes.

\subsection{Magnetic case} 

Previous discussion still leaves open question if in the absence of the electric charge one can find larger variety of NLE theories admitting the regularized black hole solutions. A well-known example is Bardeen's metric \cite{Bardeen68}, interpreted as a magnetically charged black hole solution of Einstein-NLE equations \cite{ABG00} with a ``reverse-engineered'' $\FF$-class NLE Lagrangian, which unfortunately does not respect the MWF limit. Bronnikov \cite{Bronnikov00} has noticed that a $\FF$-class NLE Lagrangian, such that the limit $\lim_{\FF\to\infty} \LL(\FF)$ exists and is finite, might admit magnetically charged solutions, regular in some sense. 

\smallskip

Some of the constraints proven in the previous section, Theorems \ref{tm:3} and \ref{tm:6}, apply to strictly magnetically charged solutions. Before we proceed with the discussion, note that one of the NLE Maxwell's equations (\ref{eq:Max2}), with $Q = 0$, may be multiplied by $\tE_r$, leading to 
\be\label{eq:aux1}
\left( \frac{P^2}{r^4} - \frac{1}{2}\,\FF \right) \LL_\FF = \frac{1}{4} \, \LL_\GG \GG
\ee
or multiplied by $\tB_r$, leading to
\be\label{eq:aux2}
\frac{1}{4} \, \LL_\FF \GG = \frac{P^2}{r^4} \, \LL_\GG \, .
\ee
Now we turn to the family of quadratic NLE Lagrangians.

\btm\label{tm:7}
Suppose that the spacetime is a static, spherically symmetric solution of the Einstein--NLE field equations with the NLE Lagrangian (\ref{eq:NLEh}), such that $h(\FF,\GG) = a\FF^2 + b\FF\GG + c\GG^2$, where $a$, $b$ and $c$ are real constants, such that the ordered pair $(b,c) \ne (0,0)$. Then, in the magnetically charged case, that is $P\ne 0$ and $Q=0$, Ricci scalar $R$ and Ricci squared $R_{ab}R^{ab}$ cannot both remain bounded as $r\to 0$.
\etm

\emph{Proof of Theorem \ref{tm:7}}. We shall divide the proof into two subcases.

\smallskip

(a) Suppose that $b = 0$. The $a = 0$ subcase is already covered by the Theorem \ref{tm:3}, so let us assume that $a \ne 0$. Using Eq.~(\ref{eq:aux2}) we have
\be
\left( \LL_\FF - \frac{8cP^2}{r^4} \right) \GG = 0 \, .
\ee
At each point where $\GG = 0$ we have $\tE_r = 0$, $\FF = 2P^2/r^4$ and
\be\label{eq:Gzero}
\LL_\FF \FF = \left( -\frac{1}{4} + \frac{4aP^2}{r^4} \right) \frac{2P^2}{r^4} \, ,
\ee
while at each point where $\GG \ne 0$ we have $\LL_\FF = 8cP^2/r^4$,
\be
\FF = \frac{1}{2a} \left( \frac{1}{4} + \frac{8cP^2}{r^4} \right)
\ee
and
\be\label{eq:Gnonzero}
\LL_\FF \FF = \left( \frac{1}{4} + \frac{8cP^2}{r^4} \right) \frac{4cP^2}{ar^4} \, .
\ee
Thus, $\LL_\FF \FF$ is a function, defined by Eq.~(\ref{eq:Gzero}) at points where $\GG = 0$ and by Eq.~(\ref{eq:Gnonzero}) at points where $\GG \ne 0$, which is unbounded as $r \to 0$, in contradiction with our basic assumptions. 

\smallskip

(b) Suppose that $b \ne 0$. First, from (\ref{eq:aux2}) we have
\be\label{eq:auxF}
\FF = \frac{r^4}{4bP^2} \, \LL_\FF\GG - \frac{2c}{b}\,\GG
\ee
that, inserted in (\ref{eq:aux1}), leads to
\begin{align}
\big( 4P^2(b^2 - 4ac) & - b r^4 \LL_\GG \big) \GG = \nonumber\\
 & = b P^2 + 2r^4 \big( b\LL_\FF\FF - a \LL_\FF\GG \big) \, .
\end{align}
If $b^2 \ne 4ac$, then it follows that $\GG$ is bounded as $r \to 0$ and, via Eq.~(\ref{eq:auxF}), the same holds for $\FF$, leading to a contradiction. On the other hand, if $b^2 = 4ac$ (which immediately excludes $c = 0$), then we have a special relation
\be
\LL_\FF = -\frac{1}{4} + \frac{b}{2c}\,\LL_\GG \ .
\ee
This implies that $\LL_\FF$ is bounded as $r \to 0$ and, as
\be
\LL_\FF \GG = \left( -\frac{1}{4} + \frac{b}{2c}\,\LL_\GG \right) \GG \, ,
\ee
the same holds for $\GG$. Furthermore, the relation $b\FF = \LL_\GG - 2c\GG$ implies that $\FF$ is bounded as $r \to 0$, leading again to a contradiction. \qed

\smallskip

The case not covered by the Theorem \ref{tm:7} above is a quadratic $\FF$-class NLE Lagrangian with $h(\FF) = a\FF^2$. Looking at the Eq.~(\ref{eq:aux2}), we see that one basic option is to take $\GG = 0$ (which again leads to a contradiction as in the proof above), while the other is to demand $\LL_\FF = 0$, that is $\FF = 1/(8a)$. In the latter case the NLE Maxwell's equations are automatically satisfied, with
\be
\tE_r^2 = \frac{P^2}{r^4} - \frac{1}{16a} \qqd \tB_r = \frac{P}{r^2} \, ,
\ee
while Einstein's field equation is reduced to
\be\label{eq:Elambda}
R_{ab} - \frac{1}{2} \, R g_{ab} + \lambda g_{ab} = 0 \, ,
\ee
with the ``effective'' cosmological constant
\be
\lambda \defeq \Lambda + \frac{1}{32 a} \, .
\ee
For each solution of this equation we have $R = 4\lambda$ and $R_{ab} R^{ab} = 4\lambda^2$, both of which are constant, thus trivially bounded. Nevertheless, the static, spherically symmetric black hole solution of the Eq.~(\ref{eq:Elambda}) is just Schwarzschild--(anti-)de Sitter black hole (see, e.g., \cite{BGH84,MuAY15} and references therein), with unbounded Kretschmann scalar $R_{abcd} R^{abcd}$. A curious feature of this solution is that the electromagnetic field is just disguised as a (part of) cosmological constant, with its imprint in a form of a nonvanishing magnetic charge $P$. In principle, one could try to glue $\GG = 0$ solution to the $\LL_\FF = 0$ solution along the $r^4 = 16aP^2$ hypersurface, but such ``chimera'' will again suffer from the same irregularities as the elementary solutions.

\subsection{Neutral case} 

In order to complete our survey, we turn finally to the neutral case, in which $Q = 0 = P$. Maxwell's equation (\ref{eq:Max1}) immediately implies $\tB_r = 0$, while (\ref{eq:Max2}) is reduced to $\LL_\FF \tE_r = 0$. Thus, at each point we have either $\tE_r = 0$, a trivial field, or $\LL_\FF = 0$. In the latter case the NLE energy-momentum tensor attains a form of the cosmological constant term $T_{ab} = (T/4) g_{ab}$ and we are again led to the special case discussed at the end of the previous subsection. It is worth taking a notice that for most of the NLE Lagrangians discussed in the literature, function $\LL_\FF$ does not have zeros \cite{ISm17}, power-Maxwell being one of the exceptions.

\section{Final remarks} 

Our results reveal severe obstructions to the prospect of black hole regularization with NLE fields and are, in some sense, complemented by the recent no-go results \cite{BJS22} for stationary, asymptotically flat, everywhere regular solutions of Einstein--NLE field equations. A fundamental obstacle is given already with the Theorem \ref{tm:1}: Electrically charged black holes in a theory with a MWF limit obeying NLE Lagrangian cannot be regular, not even in a mild sense used in this paper. In a pursue of a NLE-regularizing theory one might include magnetic charges, with a caveat that magnetic monopoles have not been observed so far. Still, even from a theoretical side, this pursuit will be limited by several constraints proven in Theorems \ref{tm:2}--\ref{tm:7}. For example, arguably the simplest type of NLE Lagrangians are quadratic ones, appearing in a weak field limit of quantum gauge theories (most important example being Euler--Heisenberg Lagrangian). However, Theorems \ref{tm:1}, \ref{tm:5} and \ref{tm:7} completely eliminate this subclass of NLE Lagrangians as candidates for regularization of the black hole singularities, in any combination of electric and magnetic charges. Born--Infeld and ModMax theories have been treated separately in Theorem \ref{tm:6}, leading to the same conclusions.

\smallskip

Some regular, magnetically charged, static black hole solutions have been found with \emph{ad hoc} proposed $\FF$-class NLE Lagrangians which, unfortunately, lack any clear physical motivation. The other strategy used in construction of regular black holes consists of evaluation of the energy-momentum tensor for a chosen metric and reconstruction of an associated NLE Lagrangian, albeit written in a coordinate form (see, e.g., \cite{TSA17, *RJ17}), rather than as functional of the electromagnetic invariants. On the other hand, recently proposed regular back hole spacetimes \cite{SV18,FLMSV21} admit an interpretation \cite{BW22} as a solution in theory with an explicit Lagrangian containing NLE and scalar fields.

\smallskip

Here we may emphasize several directions of further inquiry, motivated by the following questions. First of all, it is not clear to what extent can the constraints obtained in Theorems \ref{tm:2}--\ref{tm:6}, dealing with the dyonic case, be generalized for larger family of $\FF\GG$-class Lagrangians. One step further is to generalize the theorems from the paper when the Einstein--Hilbert action is replaced by some modified gravitational action. For example, given that one shifts to $f(R)$ class of gravitational theories \cite{RJMZ16,RFJM16,NO17}, we need to add regularity assumptions on higher derivative curvature invariants. Even more broadly, we need to investigate generalizations for the theories with the electromagnetic field nonminimally coupled to the gravitation and/or the electromagnetic Lagrangian depending on derivatives of invariants.

\smallskip

An important aspect of the proposed regular black hole solutions, which may be used to assess their physical viability, is validity of the energy conditions. In the case of NLE fields, they are controlled by the signs of the derivative $\LL_\FF$ and the trace $T$ \cite{Plebanski70,BJS21}; for example, the null energy condition holds if and only if $\LL_\FF \le 0$, while the dominant energy condition holds if and only if both $\LL_\FF \le 0$ and $T \le 0$ hold. Unfortunately, at the level of generality considered in this paper, with a mere assumption about the boundedness of curvature scalars, it is not clear how to infer something conclusive about the sign of the aforementioned functions. Relation between $\LL_\FF$, trace $T$ and derivatives of the metric function $f$, provided by the Einstein's field equation, could be utilized given that one imposes, for example, additional assumptions about convexity of the function $f$, but we are then confronted with a delicate dilemma which choice of such assumption would be ``appropriate'' in this context. Thus, the question which classes of NLE theories admit solutions with bounded curvature scalars and satisfied (some or all) energy conditions remains open.

\smallskip

Another type of spacetime singularities are those appearing at the initial or the final region of the Universe. Again, it is possible to obtain singularity-free Friedmann--Robertson--Walker cosmological solution, coupled to the NLE theory, as shown in \cite{NGSB, OL18} with $\FF$-class Lagrangians or in \cite{LK02, CG04} with Lagrangians similar to Euler--Heisenberg's. Besides FRW cosmology, it was shown that anisotropic Bianchi spaces sourced by Born--Infeld Lagrangian do not contain any singularities \cite{SB05}. It seems that at least in the cosmological context NLE-induced regularizations have more perspective, but it is not quite clear what are the general constraints delimiting such proposals.

\begin{acknowledgments}
The research was supported by the Croatian Science Foundation Project No.~IP-2020-02-9614.
\end{acknowledgments}

\bibliography{sing} 

\end{document}